\documentclass[amsfonts,aps,nofootinbib,preprint]{revtex4}
\def\S2{\bar{S}}

\def\t{\tau}

\def\sn2d{\Sn2^\dagger}

\def\({\left(}
\def\){\right)}
\def\<{\left\langle}
\def\>{\right\rangle}

\begin{document}

\title{PP-Wave Light-Cone  Free String Field Theory at Finite
Temperature}

\author{M. C. B Abdalla{\footnote{mabdalla@ift.unesp.br}},
A. L. Gadelha{\footnote{gadelha@ift.unesp.br}} and Daniel
L. Nedel{\footnote{daniel@ift.unesp.br}}}

\affiliation{Instituto de F\'{\i}sica Te\'{o}rica, Unesp, Pamplona
145, S\~{a}o Paulo, SP, 01405-900, Brazil }

\begin{abstract}
In this paper, a real-time formulation of light-cone
pp-wave string field theory at finite temperature is presented.
This is achieved  by developing the thermo field dynamics (TFD)
formalism in a second quantized string scenario. The equilibrirum
thermodynamic quantities for a pp-wave ideal string gas are derived
directly from expectation values on the second quantized string
thermal vacuum. Also, we derive the  real-time thermal pp-wave
closed string propagator. In the flat space limit it is shown that
this propagator can be  written in terms of Theta functions, exactly
as the zero temperature one. At the end, we show how supestrings
interactions can be introduced, making this approach suitable to
study the BMN dictionary at finite temperature.  
\end{abstract}

\maketitle

\section{Introduction}

The formulation of superstring theory at finite temperature is
fundamental to many applications, such as early string
cosmology and black hole physics. One of the fascinating 
features of string theory at finite temperature is the
exponential growth of states as function of energy. Such
behavior leads to a temperature above which the
partition function diverges (the Hagedorn temperature). If the
Hagedorn temperature denotes a phase transition, the true
degrees of freedom of the theory at high energy may be others
than those of the perturbative string. However, in spite of
many works about finite temperature string
theory\footnote{See for example \cite{classicos} and references
therein.}, a precise understanding of the Hagedorn temperature
and the true degrees of freedom at higher temperatures is still
lacking. In this way, beyond the applications, finite temperature
studies can provide fundamental issues of string theory itself.

Recently, the study of the finite temperature string theory
in a pp-wave background has provided great advances in this
direction 
\cite{Greene,zayas,semen,sugawara,park,bigazzi,ppwaves,POS}.
In the context of the plane wave limit of the ADS/CFT
correspondence, the BMN correspondence \cite{BMN,plefka,sadri},
it is possible to assert a dual
description of the Hagedorn temperature in terms of a
Yang-Mills temperature \cite{Greene}. In this scenario 
the Hagedorn behavior can be related to
deconfinement/confinement phase transition on the gauge side
\cite{Br}, \cite{Sud}. However, in those works superstring
interactions were neglected. As shown in \cite{Deo1,Deo2,Deo3},
the energy density of states at high temperature favours the
formation of a single long string which carries most of
the available energy. In the thermodynamic limit this long
string has numerous opportunities to intersect itself
\cite{Lowe}. So, the physical picture of an ideal string
gas at high temperature is suspect and it is fundamental
to have a finite temperature description of the superstring
interactions, in order to clarify what is the physics at the
Hagedorn limit and keep the ideal gas approximation under control.

The recent development of the plane wave limit of the ADS/CFT
correspondence brought to light an ancient string theory
subject: the Light-Cone String Field Theory (LCSFT)
\cite{kakuki1,kakuki2,gervais,greensb}.
The development of the LCSFT to study pp-wave superstring
is fundamental to understand the BMN dictionary, because
it naturally connects both sides of the correspondence when
higher genus corrections in the plane wave spectrum are taken
into account. Considering that the LCSFT is
the main tool to understand superstring interactions in a
pp-wave background \cite{SPRADLIN1,SPRADLIN2,ARY1,ARY2},
a finite temperature description of this theory could help
to understand many aspects of string thermodynamics that
still puzzle us. In general, as LCSFT is a multi-strings
theory, its finite temperature version may be the natural
scenario to study string thermodynamics in the light-cone.

In order to construct the pp-wave LCSFT at finite temperature
it is interesting to use a formalism that keeps intact all the
operator machinery of the theory at zero temperature. Therefore,
in this work we have further developed the Thermo Field
Dynamics (TFD) \cite{UME} to introduce temperature in LCSFT.
There are many characteristics of TFD that make this formalism
suitable to study string field theory at finite temperature.
Let's list some of them.
\begin{itemize}

\item
TFD is a canonical and an operator approach. The statistical
average of an observable is derived from an expectation value
in a pure state (the thermal vacuum), that depends on
temperature. In this way,  the usual quantum mechanical
perturbation theory that is used in LCSFT can be used at
finite temperature by means of this formalism. So, by
developing the TFD in the LCSFT context, we get a well-suited
tool to study string interactions at finite temperature.

\item
One of the TFD main objects is the entropy
operator. This operator has been used, for instance, to calculate
entanglement entropy of black holes \cite{Iorio}, and it can be
a powerful tool to study black hole entropy from the string point
of view.

\item
TFD has a BRST formulation \cite{OJIMA}. Then, it can be
developed further to construct a covariant string field theory
at finite temperature. In particular, in \cite{YLE1} the TFD
was used for this purpose, but it was employed within a path
integral formulation, different from the canonical operator
approach used here, as well as in \cite{OJIMA}.

\item
TFD shares common characteristics with the $C^{*}$ algebra
approach to statistical mechanics \cite{Haag, Ench}. In fact,
it was shown by Ojima \cite{OJIMA} the equivalence of TFD with
the Haag-Hugenholtz-Winnink (HHW) algebraic formulation.
Therefore, TFD can bring interesting mathematical questions
involving axiomatic statistical mechanics to the string
field theory scenario.

\item TFD is a real-time formalism: it can be used
to study time dependent processes as dissipations and
time-dependent backgrounds. Owing to the evolution of particle
distribution, a real-time formalism seems to be the appropriated
one for theories containing gravity \cite{matur}.

\end{itemize}

In this work we construct all the TFD ingredients in
the LCSFT context, in order to have tools to explore
multi-string effects at finite temperature, which include
superstring interactions, multi-string bound states formation
as well as the BMN dictionary at finite temperature.
We concentrate here on the free theory and the interactions
shall be introduced in a future work. The finite temperature formulation
of LCSFT presented here is quite general, allowing to take into account
time dependent process and out of equilibrium
thermodynamics effects. We construct the second quantized thermal vacuum for
pp-wave string theory and show that this state is an entanglement of 
strings. In the equilibrium, we compute thermodynamic quantities
of an ideal string gas directly from  expectation values on the
second quantized string thermal vacuum.
Also, we derive the real-time thermal propagator for pp-wave closed string.
We show that in the flat space limit, the thermal
propagator can be written in terms of Theta functions,
exactly as the zero temperature one.

This work is divided as follows. In section \ref{FSFT} we
introduce the basic elements of LCSFT necessary to apply
TFD. In section \ref{FTS} the system is led to finite
temperature and the TFD approach is applied in the context
of free LCSFT. The entropy operator is introduced and the
entropy as well as the thermal energy are obtained. Closing
the section we obtain the light-cone thermal distribution
for a pp-wave string gas and derive the free energy. 
Section \ref{thermalprop} is deserved to derive the pp-wave thermal
string propagator in a flat and in a pp-wave background. 
At the end, in the section \ref{concdiss} 
we discuss the relationship between first and second quantization of
string at finite temperature in the TFD approach.
The effect of the thermal Bogoliubov transformation
in both cases and the respective topological interpretation
are presented. Also we show how to introduce string 
interactions using TFD quantum mechanical perturbation theory.

\section{Free String Field Theory}
\label{FSFT}

In this section the basic concepts of the free light-cone SFT
in a pp-wave background are introduced. Following
reference \cite{SPRADLIN1}, the bosonic part of the light-cone
action for first quantized closed pp-wave superstring is:
\begin{equation}
S=\frac{e(p^{+})}{8\pi}\int d\tau\int_{0}^{4\pi|p^{+}|}
d\sigma
\left(\partial_{+}x^{i}\partial_{-}x^{j} 
- \mu^2x^{i}x^{j}\right)\delta_{ij},
\label{action1}
\end{equation}
where we have set $\alpha'=2$, $e(p^{+})$ is the signal of
$p^{+}$, $\partial_{\pm}=\partial_{\tau} \pm \partial_{\sigma}$
and $i,j=0,...\,,D-2$ labels the transverse directions in the
light-cone. The mode expansions for the string coordinate
$x^{i}(\sigma)$ and density momentum $p^{i}(\sigma)$ are:
\begin{eqnarray}
x^{i}(\sigma)&=&
x_0^{i} +
\frac{1}{\sqrt{2}}\sum_{k \neq 0}
\left(x_{|k|}^{i} - ie(k)x_{-|k|}^{i}\right )
e^{i \frac{k\sigma}{2|p^{+}|}}
\label{xmodes}
\\
p^{i}(\sigma) &=& 
\frac{1}{4\pi p^{+}}
\left[p_0^{i} +\sum_{k \neq 0}
\left(p_{|k|}^{i} - ie(k)p_{-|k|}^{i}\right )
e^{i \frac{k\sigma}{2|p^{+}|}}\right],
\label{pmodes}
\end{eqnarray}
where
\begin{equation}
x_{k}^{i} - ix_{-k}^{i}=
\sqrt{\frac{2}{\omega_{k}}}\left(\bar{\alpha}^{i}_{k} 
+ \alpha_{k}^{i\,\dagger}\right), 
\qquad
p_{k}^{i} - ie(k)p_{-k}^{i}=
\sqrt{\frac{2}{\omega_{k}}}
\left(\bar{\alpha}^{i}_{k} - \alpha_{k}^{i\,\dagger}\right),
\end{equation}
for
\begin{equation}
\omega_{k}=\sqrt{k^{2}+(2p^{+}\mu)^2}.
\end{equation}
Using $x^{i}_{k}$ and $p^{i}_{k}$ the world-sheet
Hamiltonian can be written as:
\begin{equation}
h= \frac{1}{2p^{+}}\
\sum_{k=-\infty}^{+\infty}
\left[p_k^{2} +\frac{1}{4}\omega_k^{2}x_k^{2}\right]. 
\end{equation}
As usual in the light-cone SFT these modes allow us to define
a new creation-annihilation basis, whose indices range from
$-\infty$ to $+\infty$ :
\begin{equation}
a^{i}_{k} =\frac{1}{\sqrt{\omega_{k}}}\,p^{i}_{k}
- \frac{i}{2}\frac{1}{\sqrt{\omega_{k}}}\,x^{i}_{k},
\qquad
a^{i\,\dagger}_{k} = \frac{1}{\sqrt{\omega_{k}}}\,p^{i}_{k}
+\frac{i}{2}\frac{1}{\sqrt{\omega_{k}}}\,x^{i}_{k}, 
\end{equation}  
with
$\left[a^{i}_{k}, a_{m}^{j\, \dagger}\right] 
=\delta^{ij} \delta_{km}$.
In this new basis the Hamiltonian is:
\begin{equation}
h=\frac{1}{2p {+}}\left[\sum^{\infty}_{k=-\infty}
\omega_{k}\,a^{i\,\dagger}_{k}a^{i} + A\right],
\label{fqh}
\end{equation}
where $A$ is the normal ordering constant:
\begin{equation}
A=\frac{\delta^{i}_{\,\,\,i}}{2}
\sum^{\infty}_{k=-\infty}\omega_{k}.
\label{norc}
\end{equation}
The first quantized Fock space is constructed by acting with the
creation operators $a^{i\,\dagger}_{k}$ on the vacuum 
$\left|0\right >$, which is annihilated by $a^i_k$. As we are working
with closed strings, the physical states satisfy the level matching
condition:
\begin{equation}
\sum_{i=1}^{d-2}\sum_{k=-\infty}^{\infty}k\, n^i_{k}
\left|\left.\psi\right.\right\rangle=0,
\label{lmc}
\end{equation}
where $n^i_{k}$ is the eigenvalue of the number operator 
$N_k^{i}=a^{i\,\dagger}_{k}a^{i}_{k}$.  

Let us now  define the fundamental object of the 
LCSFT, the string field $\Phi[x^{i}(\sigma),x^-, x^+]$.
The string field is a functional, associating
a collection of real numbers with each curve $x^{i}(\sigma)$
in the transverse light-cone space. It is an operator defined
in the multi-string Hilbert space
${\cal H}=
\left|{\mbox vacuum}\right\rangle\oplus_{m=1}^{\infty}{\cal H}_{m}$,
where the $m$ string Hilbert space, ${\cal H}_{m}$, is a direct
product of $m$ single Hilbert spaces ${\cal H}_{1}$. The string
field also satisfies the following equal time commutation
relation
\begin{equation}  
\left[\Phi[x^{i}(\sigma),x^{-},x^{+}],
\Phi^{\dagger}[y^{i}(\sigma),y^{-}, x^{+}]\right]
=\delta(x^{-}-y^{-})\delta^{d-2}(x^{i}-y^{i}).
\label{sqccr}
\end{equation}

As usual in light-cone field theories, the dynamics is
dictated by the Schr\"{o}dinger equation. For the string field,
it is a functional Schr\"{o}dinger equation, defined in the
light-cone configuration space by:
\begin{equation}
i\partial_{+}\Phi[x^{i}(\sigma),x^{-},x^{+}]
=\frac{1}{2}\sum_{k=-\infty}^{\infty}
\left[-\frac{\partial^2}{\partial x_{k}^{i^{2}}}
+\omega_k^2 (x_k^i)^2\right] \Phi[x^{i}(\sigma),x^{-},x^{+}]. 
\end{equation}
The solution of this equation will be useful in
section \ref{thermalprop} to derive the real-time thermal
propagator in configuration space. It is given by
\begin{equation}
\Phi\left[x^i(\sigma),x^{-}, x^{+}\right]
=\int dp^{+}\sum_{\{n_k^{i}\}}
A_{\{n^{(i)}_{k}\}}(p^{+})e^{-i\left(x^{+}p^- + x^{-}p^+\right)}
\prod_{k=-\infty}^{\infty}
\phi_{\{n^{(i)}_{k}\}}\left(x_k^i\right) + h.c.,
\label{sol}
\end{equation}
where  $\phi_{\{n^{(i)}_{k}\}}\left(x_k^i\right)$ is
written in terms of Hermite polynomials:
\begin{equation}
\phi_{ \{n^{(i)}_{k}\}}\left(x_{k}^{i}\right)
=\prod_{i=1}^{D-2}
H_{\{n_{k}^{i}\}}(\sqrt{\omega_{k}}x_{k}^{i})
e^{-\omega_{k}\frac{(x_{k}^{i})^2}{2}}
\sqrt{\frac{\sqrt{\omega_{k}}}{\sqrt{\pi}2^{n_{k}^{i}}(n_{k}^{i}!)}}.
\label{sol2}
\end{equation}
The sum over $\{n^{(i)}_{k}\}$ means the sum
over all possible sets of occupation numbers taking into account
the level matching condition (\ref{lmc}).
$A_{\{n^{(i)}_{k}\}}(p^{+})$ creates and destroys an entire
string with occupation indices $\{n^{(i)}_{k}\}$ and fixed
$p^{+}$. Thus, $A_{\{n^{(i)}_{k}\}}(p^{+})$ is the operator
that leads the $m$-string Hilbert space ${\cal H}_{m}$ into 
${\cal H}_{m\pm 1}$ with plus sign when $p^{+}<0$ and the
minus one when $p^{+}>0$. Writing
$A^{\dagger}_{\{m^{(j)}_{k}\}}(p^{+})=
A_{\{n^{(i)}_{k}\}}(-p^{+})$ for $p^{+}>0$, the commutation
relation for the second quantized creation and annihilation
operators is
\begin{equation}
\left[A_{\{n^{(i)}_{k}\}}(p^{+}),
A^{\dagger}_{\{m^{(j)}_{k}\}}(q^{+})\right] 
=\delta(p^{+}-q^{+})\,
\delta_{\{n^{(i)}_{k}\},\{m^{(j)}_{k}\}},
\label{acr}
\end{equation} 
and the vacuum of the second quantized vector space is
defined as usual:
\begin{equation}
A_{\{n^{(i)}_{k}\}}(p^{+})\left.\left|0\right.\right) =0.
\end{equation}

It is convenient for our purpose to work in the $p^+$ momentum
space and expand the string field $\Phi(p^{+})$ in terms of the
eigenstates of the string world-sheet number operators,
writing \cite{SPRADLINLEC}
\begin{equation}
\Phi(p^{+})=\frac{1}{\sqrt{|p^{+}|}}\sum_{\{n^{(i)}_{k}\}}
\left|\left.\{n^{(i)}_{k}\}\right.,p^{+}\right\rangle
A_{\{n^{(i)}_{k}\}}(p^{+}),
\label{nexp}
\end{equation}
where the ket 
$\left|\left.\{n^{(i)}_{k}\}, p^{+}\right\rangle\right.$
is a general string state in the first quantized string space. Note that
$\Phi(x, p^{+})= \left <  x \right | \Phi(p^{+}) $.
From the above expression it is clear that the string field is
a state in the first quantized closed string Fock space,
${\mathcal F}$, and an operator in the second quantized space,
${\mathcal H}$. Let's close this section writing the second quantized
string Hamlitonian. In general, the second quantized generators are
derived from the first quantized ones, by means of the map
\begin{equation}
{\cal G}=\int dp^{+}{\cal D}p(\sigma)\,p^{+}
\Phi^{\dagger}[p^{i}(\sigma),p^{+}]\,g\,
\Phi[p^{i}(\sigma),p^{+}],
\end{equation}
which gives  the second quantized Hamiltonian:
\begin{equation}
H=\int dp^{+}\,p^{+} \Phi^{\dagger}(p^{+})\,h\,\Phi(p^{+}) 
=\int dp^{+} \sum_{\{n^{(i)}_{k}\}}E_{\{n^{(i)}_{k}\}}
A^{\dagger}_{\{n^{(i)}_{k}\}}(p^{+})A_{\{n^{(i)}_{k}\}}(p^{+}),
\label{sqh}
\end{equation}
with
\begin{equation}
E_{\{n^{(i)}_{k}\}}=
\frac{1}{2p^{+}}\left[\sum^{\infty}_{k=-\infty}w_{k}n^{i}_{k}
+ A\right],
\label{energy}
\end{equation}
where $A$ is defined in (\ref{norc}).

With the ingredients of the free light-cone string field
theory in the pp-wave background here presented, we are able
to construct its finite temperature analogous and obtain the
free string gas for this system.

\section{Finite Temperature System}
\label{FTS}

TFD, when proposed by Takahashi and Umezawa \cite{UME}, was based
on the idea of interpreting the statistical average of an operator,
${\cal O}$, as its expectation value in a temperature dependent
state called thermal vacuum:
\begin{equation}
\left\langle{\cal O}\right\rangle\equiv
\frac{{\mbox Tr}\left[{\cal O}\,e^{\beta H}\right]}
{{\mbox Tr}\left[e^{\beta H}\right]}
\equiv\left\langle\left.0\left(\beta\right)\right.\right|
{\cal O}\left|\left.0\left(\beta\right)\right.\right\rangle,
\label{tfdsp}
\end{equation}
where $\beta$ is the inverse of the temperature and $H$ represents
the Hamiltonian operator of the system under consideration.
Basically two implications lie behind such a proposal. Firstly,
it is necessary a duplication of the system's degrees of freedom.
Secondly, the use of an specific Bogoliubov transformation, the
so-called thermal Bogoliubov transformation, to lead the system
to a finite temperature.
The implementation of the doubling,
as well as the specifications for the thermal Bogoliubov
transformation, will be presented in the following subsections,
in the context of the system approached in this paper.
This section as a whole has the aim to introduce the basic elements
of TFD necessary to obtain the free light-cone SFT at finite
temperature, with the perspective of developing a TFD construction
to study string field interactions at finite temperature.
Here we derive the Bose multi-string distributions and the
free energy for the ideal string gas in the pp-wave background.
Entropy and thermal energy are also calculated from the thermal
expectation values.

\subsection{The Duplication of Degrees of Freedom}

As it was mentioned above, the TFD approach consists of a
doubling of the system's degrees of freedom followed by a
thermal Bogoliubov transformation. 
Following the TFD algorithm the duplication is implemented by
introducing a copy of the original Hilbert space of the system
${\cal H}$. Denoting the copy as $\widetilde{{\cal H}}$, the
total Hilbert space that take place for the doubled system is 
$\widehat{{\cal H}}={\cal H}\otimes\widetilde{\cal H}$.
Here and in the following, we refer to the copy of the original
system as tilde system or auxiliary system.
There is a map between both systems given by the tilde conjugation rules 
\begin{eqnarray}
\left(AB\right)\widetilde{^{{}}}
&=&\widetilde{A}\widetilde{B},
\\
\left(c_{1}A+c_{2}B\right)\widetilde{^{{}}}
&=&\left(c_{1}^{*}\widetilde{A}
+c_{2}^{*}\widetilde{B}\right),
\\
\left(A^{\dagger }\right)\widetilde{^{{}}}
&=&\widetilde{A}^{\dagger},
\\
\left(\widetilde{A}\right) \widetilde{^{{}}}&=&A,
\\
\left(\left|0\left(\theta\right)\right\rangle\right)\widetilde{^{{}}}
&=&\left|0\left(\theta \right)\right\rangle,
\\
\left(\left\langle 0\left(\beta\right)\right|\right)\widetilde{^{{}}}
&=&\left\langle 0\left( \beta \right) \right|,
\end{eqnarray}
with $A$ and $B$ representing bosonic operators and
$c_{1}$, $c_{2}\in {\mathbb C}$.

As the string field is a state in ${\cal F}$, the doubling of
the degrees of freedom implies a duplication of the basis
states. Defining the duplicated number basis by:
\begin{equation}
\left|\left.\left.\{n^{(i)}_{k}\}
\right.\right\rangle\!\!\right\rangle =
\left|\left.\{n^{(i)}_{k}\},p^{+} \right.\right\rangle \otimes
\widetilde{
\left|\left.\{n^{(i)}_{k}\}, p^{+}\right.\right\rangle},
\end{equation}
the string  and tilde string  fields have the following
expansion in the total Hilbert space:
\begin{eqnarray}
{\bf \Phi}(p^{+}) &=&
 \frac{1}{\sqrt{|p^{+}|}}\sum_{\{n^{(i)}_{k}\}}
\left|\left.\left.\{n^{(i)}_{k}\},p^{+}
\right.\right\rangle\!\!\right\rangle 
A_{\{n^{(i)}_{k}\}}(p^{+})\nonumber \\
\widetilde{{\bf \Phi}}(p^{+})&=&
\frac{1}{\sqrt{|p^{+}|}}\sum_{\{n^{(i)}_{k}\}}
\left|\left.\left.\{n^{(i)}_{k}\},p^{+}
\right.\right\rangle\!\!\right\rangle
\widetilde{A}_{\{n^{(i)}_{k}\}}(p^{+}),
\label{dnexp}
\end{eqnarray}
where the tilde field expansion is obtained just using the
tilde conjugation rules. Also, the tilde conjugation rules
provide that the dynamics of the auxiliar system is the same
as the original one. At this point it is interesting to
 emphasize that the aplication of the TFD alogorithm in second
quantized string demands a duplication of both, the first
quantized Hilbert space ${\cal F}$ and the second quantized
Hilbert space ${\cal H}$. In the string first quantized 
TFD aplications, the auxilary system is interpretated as an
auxiliary string \cite{TORUS}. So, the field string is now a
functional of two independent strings. This fact gives rise
to an interesting topological interpretation, to be
discussed in the last section.    

The canonical commutation relations for the tilde string
creation and annihilation operators are the same as the
original ones given at (\ref{acr}), and elements from different
spaces commute. The vacuum of the total system is now defined
by
\begin{equation}
A_{\{n^{(i)}_{k}\}}(p^{+})
\left.\left.\left|0\right.\right)\!\right)=
\widetilde{A}_{\{n^{(i)}_{k}\}}(p^{+})
\left.\left.\left|0\right.\right)\!\right)=0.
\label{dsqv}
\end{equation}

Finally, let's define the the total second quantized Hamiltonian.
The Heisenberg equations for the Heisenberg original and tilde fields,
together with the tilde conjugation rules, demand that the total
second quantized Hamiltonian is not $H$ anymore, but it is given by
\begin{equation}
\widehat{H}=H-\widetilde{H},
\label{tsqh}
\end{equation}
where $H$ is defined in (\ref{sqh}) and $\widetilde{H}$ is
obtained by using the  tilde conjugation rules. This Hamiltonian
structure is usual in the TFD approach and it is very important
to define an interaction representation at finite temperature,
which allows to use the same quantum mechanical perturbation
theory defined at zero temperature.
\subsection{Thermal Vacuum and Thermal Operators}

Once we have defined the doubled system, we are ready to
introduce the thermal Bogoliubov transformation.
It is such that its generator,
{\bf G}, consists of bilinear terms composed by creation and
annihilation operators from both, original and tilde systems.
Besides this, it must satisfy what Umezawa called
${\bf G}$-symmetry of TFD \cite{ume93}, defined by the
following three requirements, generalized here to
string field theory: the generator must induce a
transformation of the form
\begin{eqnarray}
\left(
\begin{array}{c}
A_{\{n^{(i)}_{k}\}}(\theta) \\
\widetilde{A}^{\dagger}_{\{n^{(i)}_{k}\}}(\theta)
\end{array}
\right) &=&e^{-i{\bf G}}\left(
\begin{array}{c}
A_{\{n^{(i)}_{k}\}} \\
\widetilde{A}^{\dagger }_{\{n^{(i)}_{k}\}}
\end{array}
\right) e^{i{\bf G}}= {\mathbb B}_{\{n^{(i)}_{k}\}}\left(
\begin{array}{c}
A_{\{n^{(i)}_{k}\}} \\
\widetilde{A}^{\dagger }_{\{n^{(i)}_{k}\}}
\end{array}
\right) , \label{tbt} \\
\left(\begin{array}{cc}
A^{\dagger}_{\{n^{(i)}_{k}\}}(\theta) &
-\widetilde{A}_{\{n^{(i)}_{k}\}}(\theta)
\end{array}
\right) &=&\left(
\begin{array}{cc}
A^{\dagger }_{\{n^{(i)}_{k}\}} &
-\widetilde{A}_{\{n^{(i)}_{k}\}}
\end{array}
\right) {\mathbb B}^{-1}_{\{n^{(i)}_{k}\}},
\label{tbti}
\end{eqnarray}
where by $\theta$ we denote the transformation parameter that
encodes the temperature and further relevant
theory parameters dependences. ${\mathbb B}$ is the transformation
matrix with the following general form
\begin{eqnarray}
{\mathbb B}_{\{n^{(i)}_{k}\}}=\left(
\begin{array}{cc}
u_{\{n^{(i)}_{k}\}} & v_{\{n^{(i)}_{k}\}} \\
v^{*}_{\{n^{(i)}_{k}\}} &
u^{*}_{\{n^{(i)}_{k}\}}
\end{array}
\right) ,
\label{tbm}
\end{eqnarray}
with
$u_{\{n^{(i)}_{k}\}}u^{*}_{\{n^{(i)}_{k}\}}-
v_{\{n^{(i)}_{k}\}}v^{*}_{\{n^{(i)}_{k}\}}=1$ for bosonic systems
in order to the transformation be canonical. The generator must
commute with the total Hamiltonian of the system (\ref{tsqh}) and
finally, the generator must change its sign under tilde
conjugation rules implying thermal vacuum
invariance under tilde conjugation.

The general TFD structure presented above allows us to observe
that more than one generator satisfy such a requirements. In
fact, TFD can be constructed taking into account a general
generator that is a linear combination of the possible choices
of $\bf{G}$
\cite{ume93,micro,elmf,henning,agvsu,agvproc,ag,agn1,agn2}.
Here we will restrict the TFD construction to
one generator only, following the originally proposed by
Takahashi and Umezawa
\begin{equation}
{\bf G}=-i\sum_{\{n^{(i)}_{k}\}}\theta_{\{n^{(i)}_{k}\}}
\left(A_{\{n^{(i)}_{k}\}}\widetilde{A}_{\{n^{(i)}_{k}\}}
-A^{\dagger}_{\{n^{(i)}_{k}\}}
\widetilde{A}^{\dagger }_{\{n^{(i)}_{k}\}}
\right).
\label{sqg}
\end{equation}
It satisfies the $\bf {G}$-symmetry and the
thermal Bogoliubov matrix elements presented in (\ref{tbm})
are explicitly written as
\begin{equation}
u_{\{n^{(i)}_{k}\}}=\cosh(\theta_{\{n^{(i)}_{k}\}}), \qquad
v_{\{n^{(i)}_{k}\}}=-\sinh(\theta_{\{n^{(i)}_{k}\}}).
\label{tbte}
\end{equation}
The second quantized thermal vacuum is defined by
\begin{equation}
A_{\{n^{(i)}_{k}\}}(\theta)
\left.\left|0(\theta)\right.\right)=
\widetilde{A}_{\{n^{(i)}_{k}\}}(\theta)
\left.\left|0(\theta)\right.\right)=0,
\label{tvd}
\end{equation}
where the thermal annihilation operators were obtained from
expression (\ref{tbti}).
The second quantized thermal
space is constructed by cyclic applications of the thermal
string creation operators $A_{\{n^{(i)}_{k}\}}^{\dagger}(\theta)$
and $\widetilde{A}_{\{n^{(i)}_{k}\}}^{\dagger}(\theta)$.
The above definition for the thermal vacuum gives rise
to the so-called thermal state condition
\begin{equation}
\left[A_{\{n^{(i)}_{k}\}}- \tanh(\theta_{\{n^{(i)}_{k}\}})
\widetilde{A}^{\dagger}_{\{n^{(i)}_{k}\}}\right]
\left.\left|0(\theta)\right.\right)=0.
\label{sqtsc}
\end{equation}
From (\ref{sqtsc}) it is possible to derive the KMS condition to be
presented later.
Also, it tells us that the annihilation operator of the
original system does not annihilate the transformed vacuum.
In fact, the thermal vacuum satisfying expression
(\ref{tvd}) is explicitly written as
\begin{equation}
\left.\left|0(\theta)\right.\right)=
e^{-i{\bf G}}\left.\left.\left|0\right.\right)\!\right)=
\prod_{\{n^{(i)}_{k}\}}
\left(\frac{1}{\cosh(\theta_{\{n^{(i)}_{k}\}})}\right)
e^{\tanh(\theta_{\{n^{(i)}_{k}\}})
A^{\dagger}_{\{n^{(i)}_{k}\}}\(p^{+}\)
\widetilde{A}^{\dagger}_{\{n^{(i)}_{k}\}}\(p^{+}\)}
\left.\left.\left|0\right.\right)\!\right),
\label{esqtv}
\end{equation}
with doubled vacuum  defined in (\ref{dsqv}).
The above structure manifests the thermal vacuum
invariance under tilde conjugation and
shows us that the second quantized thermal
vacuum is a condensed state of string thermal pairs
$A^{\dagger}_{\{n^{(i)}_{k}\}}\(p^{+}\)
\widetilde{A}^{\dagger}_{\{n^{(i)}_{k}\}}\(p^{+}\)$.

\subsection{The Entropy Operator}

The aim of this section is to introduce the TFD entropy
operator presenting some properties of it.
The name of such an operator will be clear at the end of
this subsection.

In our extension of TFD for string field theory
the second quantized entropy operator is defined as
\begin{equation}
\widehat{K}=K-\widetilde{K},
\end{equation}
where
\begin{equation}
K=-\sum_{\{n^{(i)}_{k}\}}
\left[A^{\dagger}_{\{n^{(i)}_{k}\}}(p^{+})
A_{\{n^{(i)}_{k}\}}(p^{+})
\ln\sinh^{2}(\theta_{\{n^{(i)}_{k}\}})
-A_{\{n^{(i)}_{k}\}}(p^{+})
A^{\dagger}_{\{n^{(i)}_{k}\}}(p^{+})
\ln\cosh^{2}(\theta_{\{n^{(i)}_{k}\}})\right],
\label{kb}
\end{equation}
and $\widetilde{K}$ is obtained from $K$
by tilde conjugation rules. The operator $\widehat{K}$
commutes with the generator ${\bf G}$ and  the total
Hamiltonian $\widehat{H}$.
An interesting property of the entropy operator is the fact
that it  can be used to lead the system from zero to a
finite temperature, by considering the following expression
for the thermal vacuum
\begin{equation}
\left.\left|0(\theta)\right.\right) =
e^{-\frac{1}{2}K}
e^{\sum\limits_{\{n^{(i)}_{k}\}}
A_{\{n^{(i)}_{k}\}}^{\dagger }\(p^{+}\)
\widetilde{A}^{\dagger}_{\{n^{(i)}_{k}\}}\(p^{+}\)}
\left.\left.\left|0\right.\right)\!\right).
\label{tsqvk}
\end{equation}
Using the relations
\begin{eqnarray}
e^{-\frac{1}{2} K}\left.\left.\left|0\right.\right)\!\right)
&=&\prod\limits_{\{n^{(i)}_{k}\}}
\left( \frac{1}{\cosh(\theta_{\{n^{(i)}_{k}\}})}\right)
\left.\left.\left|0\right.\right)\!\right),
\\
e^{-\frac{1}{2}K}
A^{\dagger}_{\{n^{(i)}_{k}\}}\(p^{+}\)
e^{\frac{1}{2} K}&=&
\tanh(\theta_{\{n^{(i)}_{k}\}})
A_{\{n^{(i)}_{k}\}}^{\dagger }\(p^{+}\),
\end{eqnarray}
one finds the same thermal vacuum obtained from the thermal
Bogoliubov generator, whose structure is presented
in (\ref{esqtv}). $\widetilde{K}$ also furnishes the same
thermal vacuum, making use of the
tilde conjugation version of the above expressions.

One can go further in the use of the entropy operator to
generate the thermal vacuum: if we define the following entangled state
\begin{eqnarray}
\left|\left.{\cal I}\right.\right)
&\equiv&
e^{\sum\limits_{\{n^{(i)}_{k}\}}
A_{\{n^{(i)}_{k}\}}^{\dagger }\(p^{+}\)
\widetilde{A}^{\dagger}_{\{n^{(i)}_{k}\}}\(p^{+}\)}
\left.\left.\left|0\right.\right)\!\right)
\\
&=&\sum\limits_{{\bf\{n^{(i)}_{k}\}}}
\left.\left.\left|{\bf\{n^{(i)}_{k}\}}\right.\right)\!\right)
\\
&=&\left.\left.\left|0\right.\right)\!\right)
+\sum\limits_{\{n^{(i)}_{k}\}}
\left.\left.\left|\{n^{(i)}_{k}\}\right.\right)\!\right)
+\sum\limits_{\{n^{(i)}_{k}\}}\sum\limits_{\{m^{(i)}_{k}\}}
\left.\left.\left|\{n^{(i)}_{k}\},\{m^{(i)}_{k}\}
\right.\right)\!\right)
\\
&+&\sum\limits_{\{n^{(i)}_{k}\}}\sum\limits_{\{m^{(i)}_{k}\}}
\sum\limits_{\{l^{(i)}_{k}\}}
\left.\left.\left|
\{n^{(i)}_{k}\},\{m^{(i)}_{k}\},\{l^{(i)}_{k}\}
\right.\right)\!\right)+...\,\,,
\label{ids}
\end{eqnarray}
the expression (\ref{tsqvk}) can be written as
\begin{equation}
\left.\left|0(\theta)\right.\right)=
e^{-\frac{1}{2}K}
\left|\left.{\cal I}\right.\right),
\end{equation}
and by explicit calculation one finds
\begin{equation}
\left.\left|0(\theta)\right.\right)
=\sum_{\{n^{(i)}_{k}\}}\sqrt{W_{\{n^{(i)}_{k}\}}}
\left|\left.{\cal I}\right.\right),
\label{etve}
\end{equation}
where
\begin{equation}
W_{\{n^{(i)}_{k}\}}=\prod_{i=1}^{D-2}\prod_{k=-\infty}^{\infty}
\frac{\sinh^{2n^{i}_{k}}(\theta_{\{n^{i}_{k}\}})}
{\cosh^{2(n^{i}_{k}+1)}(\theta_{\{n^{i }_{k}\}})},
\label{wcoef}
\end{equation}
is related with the density matrix of the system.

The next property of the entropy operator is exactly
what justifies its name: the thermal expectation value of
the operator $K$ defined in (\ref{kb});
\begin{equation}
{\cal S}=\int dp^{+}
\left(\left.0(\theta)\right|\right.
K
\left.\left|0(\theta)\right.\right).
\end{equation}
The strategy is to use the inverse of thermal Bogoliubov
transformation given by (\ref{tbt}) and (\ref{tbti}) with
matrix elements presented in (\ref{tbte}), to write the
second quantized creation and annihilation operators in
terms of thermal operators. Applying them to the thermal
vacuum defined in (\ref{tvd}), results the following
expression
\begin{equation}
{\cal S}=-\int dp^{+}\sum_{\{n^{(i)}_{k}\}}
\left[{\cal N}_{\{n^{(i)}_{k}\}}(\theta)
\ln\({\cal N}_{\{n^{(i)}_{k}\}}(\theta)\)
-\(1+{\cal N}_{\{n^{(i)}_{k}\}}(\theta)\)
\ln\(1+{\cal N}_{\{n^{(i)}_{k}\}}(\theta)\)\right],
\label{entropy}
\end{equation}
where ${\cal N}_{\{n^{(i)}_{k}\}}(\theta)$ is the thermal expectation
value of the second quantized number operator
\begin{equation}
{\cal N}_{\{n^{(i)}_{k}\}}(\theta)
=\left.\left( 0(\theta)\right.\right|
N(p^{+})
\left.\left|0(\theta)\right.\right)
=\sinh^{2}(\theta_{\{n^{(i)}_{k}\}}),
\end{equation}
The $p^{+}$ dependence in (\ref{entropy}) is
implicit in the parameter $\theta$
as it will be clear in the next subsection where it will be shown
that the expectation value of the second quantized number
operator is in fact the thermal distribution of the system.
Note that the expression (\ref{entropy}) is formally identical
to the general expression for the entropy of bosonic systems,
explaining why the $K$ operator is called the entropy operator.

One can proceed in a different way and perform the thermal
expectation value of the entropy operator using the expression
(\ref{etve}) for the thermal vacuum and hence obtaining
the entropy of the system written in terms of the
coefficients $W_{\{n^{(i)}_{k}\}}$:
\begin{equation}
{\cal S}=-\int dp^{+}\sum_{\{n^{(i)}_{k}\}}W_{\{n^{(i)}_{k}\}}
\ln W_{\{n^{(i)}_{k}\}}.
\end{equation}

It is clear from equation (\ref{etve}) that the thermal vacuum
generated by the Bogoliubov operator or by the entropy operator
has an entanglement structure. Such an structure was explored in
black hole physics \cite{Iorio}, where the entanglement is
dynamically generated by gravitational interactions. In this context,
a TFD-like entropy operator was used to derive the entanglement
black hole entropy and the result agree with the area law.
In this scenario, it will be very interesting to use the entropy
operator constructed here to compute black hole entropy from string
entanglement.

\subsection{The Thermal Energy and The Free Energy}

As mentioned before, thermal quantities in TFD are
obtained from the thermal expectation value of dynamical
operators of the original (non tilde) system.
As we are working in the light-cone gauge, in order to obtain
the thermal energy of the system the
second quantized version of the light-cone energy
$p^{0}=\frac{1}{\sqrt{2}}\left(p^{-}+p^{+}\right)$
 is needed:
\begin{equation}
P^{0}=\int dp^{+}\,p^{+} \Phi^{\dagger}(p^{+})\,p^{0}\,\Phi(p^{+})
=\frac{1}{\sqrt{2}}
\int dp^{+} \sum_{\{n^{(i)}_{k}\}}\left(E_{\{n^{(i)}_{k}\}}
+p^{+}\right)
A^{\dagger}_{\{n^{(i)}_{k}\}}(p^{+})A_{\{n^{(i)}_{k}\}}(p^{+})
\end{equation}
where $E_{\{n^{(i)}_{k}\}}$ is defined in (\ref{energy}).
The thermal energy of the system is the
expectation value of the second quantized operator $P^{0}$.
Denoting by ${\cal E}$ this energy, one has
\begin{equation}
{\cal E}=\left(\left.0(\theta)\right|\right.
P^{0}
\left.\left|0(\theta)\right.\right) = 
\frac{1}{\sqrt{2}}\int dp^{+}\sum_{\{n^{(i)}_{k}\}}
\left(E_{\{n^{(i)}_{k}\}}+p^{+}\right)
\sinh^{2}(\theta_{\{n^{(i)}_{k}\}}).
\label{tenergy}
\end{equation}

With the expression for the entropy and thermal energy
of the second quantized system, let us note that such
expressions are given in terms of the Bogoliubov matrix
elements. Those thermal quantities we named in advance,
once the thermal distribution of the system was not
presented yet. In fact, the term thermal or temperature
we have been using all along the paper was always in
this sense. The TFD construction presented here is quite
general in order to deal with time-dependent as well as
dissipation phenomena. However our specific interest in
this work is to deal with the equilibrium situation.
Such a specification can be realized as follows.
If ${\cal E}$ is the thermal energy and ${\cal S}$ the
entropy, the free energy is defined as usual
\begin{equation}
{\cal F} = {\cal E}-T{\cal S},
\label{fenergy}
\end{equation}
where $T$ is identified as the temperature. Now, one can
find that, the differentiation of the free energy with respect
to the $\theta$ parameter is an extremum for
\begin{equation}
{\cal N}_{\{n^{(i)}_{k}\}}= \sinh^{2}(\theta_{\{n^{(i)}_{k}\}})
=\frac{e^{-\frac{\beta}{\sqrt{2}} (E_{\{n^{(i)}_{k}\}}+\,\,p^{+})}}
{1-e^{-\frac{\beta}{\sqrt{2}} (E_{\{n^{(i)}_{k}\}}+\,\,p^{+})}}.
\end{equation}
In this way, the Bose distribution for ideal string gas is
derived. Replacing the above distribution in the expressions
(\ref{entropy}) and (\ref{tenergy}), the equation
(\ref{fenergy}) for the free energy can be written as
\begin{equation}
{\cal F}=\frac{1}{\beta}\int dp^{+}\sum_{\{n^{(i)}_{k}\}}
\ln\left(1-e^{-\frac{\beta}{\sqrt{2}}(E_{\{n^{(i)}_{k}\}}
+\,\,p^{+})}\right).
\end{equation}
In this expression, the sum over $\{n_{k}^{(i)}\}$ can be performed
for each value of $k, i$,  and must be constrained in order to take into
account the level matching condition (\ref{lmc}).
This is usually made by means of a Lagrange multiplier $\tau_1$
\begin{equation}
\sum_{\{n^{(i)}_{k}\}}\rightarrow
\int_{-1/2}^{1/2} d\tau_{1}
\sum_{n^{i}_{k}}\prod_{i=1}^{D-2}\prod_{k=-\infty}^{\infty}
e^{i2\pi kn_{k}^{i}\tau_{1}}
\end{equation}
Expanding the logarithm, the result is:
\begin{equation}
{\cal F}=\sum_{r=1}^{\infty}\frac{1}{4\sqrt{2}\pi}
\int_{0}^{\infty}\frac{d\tau_{2}}{\tau_{2}^{2}}
\int_{-\frac{1}{2}}^{\frac{1}{2}} d\tau_{1}
\,e^{-\frac{r^{2}\beta^{2}}{8\pi\tau_{2}}}
\,e^{-2\pi\tau_{2}(D-2)\gamma_{0}\(m\)}
\prod_{k=-\infty}^{\infty}
\left(
\frac{1}{1-e^{2\pi\(-\tau_{2}\omega_{k}+i\tau_{1}k\)}}
\right)^{D-2},
\label{sqfree}
\end{equation}
where
\begin{equation}
\tau_{2}=\frac{r\beta}{4\sqrt{2}\pi p^{+}},
\qquad
\gamma_{0}\(m\)
=\frac{1}{2}\sum_{k=-\infty}^{\infty}\omega_{k}
=\frac{m}{2}+\sum_{k=1}^{\infty}\sqrt{k^{2}+m^{2}}.
\end{equation}
Here $m=2p^{+}\mu$ was defined in order to match up the
notation of Ref. \cite{zayas} (see also Ref. \cite{bigazzi}
for a detailed study about the zero point energy in general
pp-waves backgrounds). This expression is the light-cone
free energy for a pp-wave string gas, derived from second quantized
string theory, just by evaluating second quantized operators in a pure
state. Although this result is already known, the derivation of a
string gas from a second quantized string theory is interesting by
itself, since we have worked directly in the multi-string Hilbert
space. With the TFD formulation of a free LCSFT at finite
temperature, we hope to be able to introduce string interactions
and derive a free energy for a non-ideal string gas.
We will return to this point in the last section where a propose is
presented. For now, we are ready to deal with a necessary ingredient
to study LCSFT at finite temperature: the thermal propagator.

\section{The Thermal Propagator}
\label{thermalprop}

In order to compute amplitudes  in light-cone string 
theory at finite temperature, it is necessary to calculate
the light-cone thermal propagator. The main goal of this
section is to derive the light-cone thermal propagator
for pp-wave closed string from the TFD formulation of
LCSFT at finite temperature discussed early.  For the flat
space limit the thermal propagator can be expressed in
terms of Theta functions, exactly as it was done in the
zero temperature limit \cite{SIERRA,Tow,POLT}. Owing to the
modular properties of the Theta functions, this result
can be very useful to study the dynamics of the theory 
at the Hagedorn limit.

In general, the real-time propagator for a field $\Phi(x)$
at finite temperature has the following
matrix structure \cite{LAN}:
\begin{equation}
G_{\Phi}\left(x-y\right)\rightarrow \left( \begin{array}{cc}
G_{\Phi}^{11}\left(x-y\right)& G_{\Phi}^{12}\left(x-y\right)
\\
G_{\Phi}^{21}\left(x-y\right) & G_{\Phi}^{22}\left(x-y\right) 
\end{array} \right).
\end{equation}
The $G_{\Phi}^{11}$ component is the usual physical
propagator and it is this object that we are interested
in. The other three propagators are considered as
auxiliary ones (unphysical). They are useful to eliminate
undesirable divergences coming from delta function products
\cite{LAN}. In the Schwinger's closed-time path formalism
this structure is a consequence of the time path, that goes
from $ - \infty$ to $+\infty$ then back to $-\infty$
\cite{Sch, Rivers}. In the TFD approach this matrix
structure comes directly from the doubling of the field
variables. For a general field $\Phi$, the TFD propagators
can be easily derived in a canonical way, by evaluating
the following expectation value:
\begin{eqnarray}
G_{TFD}^{11} (x-y) &=& 
\left( 0\left(\beta\right)\right|
T\Phi(x)\Phi(y)\left|0\left(\beta\right)
\right),
\nonumber 
\\
G_{TFD}^{12} (x-y) &=& 
\left( 0\left(\beta\right)\right|
T\Phi(x)\widetilde{\Phi}(y)
\left|0\left(\beta\right)\right),
\nonumber 
\\
G_{TFD}^{21} (x-y) &=& 
\left(0\left(\beta\right)\right
|T\widetilde{\Phi}(x)\Phi(y)
\left|0\left(\beta\right)\right),
\nonumber \\
G_{TFD}^{22} (x-y)  &=& 
\left( 0\left(\beta\right)\right|
T\widetilde{\Phi}(x)
\widetilde{\Phi}(y)
\left|0\left(\beta\right)\right),
\end{eqnarray}
where $T$ means time ordering and
$\left|\left.0\left(\beta\right)\right.\right)$
is the thermal vacuum defined for this field. The
$G_{\Phi}^{11}$ component of the TFD propagator is the same
as the Schwinger's propagator, while the auxiliary ones are
related by means of a time path deformation \cite{ume93}. 

Let's go back to the closed string theory. The pp-wave thermal
physical propagator in the light-cone configuration space is
defined by:
\begin{equation}
G\left(x^i(\sigma)-y^i(\sigma),\Delta x^{-},\Delta x^{+}\right )_{\beta}
=\left\langle 0\left(\beta\right)\right|
\Phi(x^i(\sigma),x_{0}^{-}, x^{+})
\Phi(y^i(\sigma),y_{0}^{-}, y^+)
\left|0\left(\beta\right)\right\rangle, 
\label{propag}
\end{equation}
where $\Delta x^{-}= x_{0}^{-} -y_{0}^{-}$ and it is assumed
that the light-cone time interval $\Delta x^{+}$ is always
positive. In this expression it is necessary to impose the level
matching condition in the thermal expectation value. We
will come back to this point later.

In order to calculate
the propagator, we are going to use the solution of the 
Schr\"{o}dinger equation in configuration space,  given by 
(\ref{sol}) and (\ref{sol2}).
Using the expansion (\ref{sol}), the commutation relations defined
in (\ref{sqccr}) and the inverse of the thermal Bogoliubov
transformation, the expectation value (\ref{propag}) can
be written in terms of the modes $y^{i}_{k}$, $x^{i}_{k}$ :
\begin{eqnarray} 
G \left(\{x^{i}_{k}\},\{y^{i}_{k}\},
\Delta x^{-}, \Delta x^{+}\right )_{\beta} 
&=&
\sum_{n^{i}_{k}=0}^{\infty}\int \frac{dp^{+}}{2\pi}
e^{-i(\Delta x^{-}p^+ +\Delta x^{+}p^-)}
\nonumber
\\
\times\prod_{i=1}^{D-2}\prod_{k=-\infty}^{\infty}
\!\!\!\!&&\!\!\!\!
\sqrt{\frac{\omega_k}{\pi}}
\frac{ H_{n^{i}_{k}}(\sqrt{\omega_{k}}x_k^{i})
H_{n_k^{i}}(\sqrt{\omega_k}y_{k}^{i})}{2^{n_{k}^{i}}(n_k^{i}!)}
e^{-\frac{1}{2}\omega_{k}\((x^{i}_{k})^2+(y^{i}_{k})^{2}\)}
\nonumber
\\
&+&\sum_{n^{i}_{k}=0}^{\infty} \int \frac{dp^{+}}{\pi}
e^{-i(\Delta x^{-}p^+ +\Delta x^{+}p^-)}
\nonumber
\\
\times\prod_{i=1}^{D-2}\prod_{k=-\infty}^{\infty}
\!\!\!\!&&\!\!\!\!
\sqrt{\frac{\omega_{k}}{\pi}}
\frac{ H_{n^{i}_{k}}(\sqrt{\omega_{k}}x_{k}^{i})
H_{n_{k}^{i}}(\sqrt{\omega_{k}}y_{k}^{i})}{2^{n_{k}^{i}}(n_{k}^{i}!)}
e^{-\frac{1}{2}\omega_{k}\((x^{i}_{k})^2+(y^{i}_{k})^{2}\)}
\sinh^{2}(\theta_{n^{i}_{k}}).
\nonumber
\\
\label{propag2}
\end{eqnarray}
As usual in real-time formalisms, the propagator is
a sum of a zero temperature part (that is equal to
the zero temperature propagator) plus finite temperature
corrections. By using the thermal state condition
(\ref{sqtsc}) it is easy to see that this propagator
satisfies the KMS condition:
\begin{equation}
G\left(\{x^{i}_{k}\},\{y^{i}_{k}\},\Delta x^{-},
\Delta x^{+}\right)_{\beta} 
=
G\left(\{x^{i}_{k}\},\{y^{i}_{k}\},
\Delta x^{-},\Delta x^{+}+i\beta\right )_{\beta}. 
\end{equation}

Now the expression for the propagator will be simplified.
The first part is exactly the zero temperature contribution,
so let's begin with the temperature corrections, that sit in
the second part. By defining the parameter $W_{lk}$
\begin{equation}
W_{k\,l}=e^{i\pi\omega_{k}\tau_{\,l}}=q_{\,l}^{\omega_{k}},
\qquad \tau_{\,l} 
= \frac{1}{2\pi p^{+}}\(-\Delta x^{+} + i\frac{\beta l}{\sqrt{2}}\),
\label{tau}
\end{equation}
with $q_{\,l}=e^{i\pi\t_{\,l}}$, and writing the propagator as
\begin{eqnarray}
G\left(\{x^{i}_{k}\},\{y^{i}_{k}\},\Delta x^{-},\Delta x^{+}\right)_{\beta}
&=& 
G\left(\{x^{i}_{k}\},\{y^{i}_{k}\},\Delta x^{-},\Delta x^{+}\right)_{T=0}
\nonumber
\\
&+&\Delta\left(\{x^{i}_{k}\},\{y^{i}_{k}\},
\Delta x^{-},\Delta x^{+}\right)_{\beta}, 
\end{eqnarray}
the thermal corrections are given by the expression:
\begin{eqnarray}
\Delta\left(\{x^{i}_{k}\},\{y^{i}_{k}\},
\Delta x^{-},\Delta x^{+}\right )_{\beta}
= \sum_{l=1}^{\infty}\sum_{n^{i}_{k}=0}^{\infty}
\int\frac{dp^+}{\pi}
e^{-ip^+(\Delta x^{-} -i\frac{\beta l}{\sqrt{2}})}
e^{-i\frac{A}{2p^{+}}(\Delta x^{+} -i\frac{\beta l}{\sqrt{2}})}
\nonumber
\\
\times\prod_{i=1}^{D-2}\prod_{k=-\infty}^{\infty}
\sqrt{\frac{\omega_k}{\pi}}
\frac{H_{n^{i}_{k}}(\sqrt{\omega_{k}}x_{k}^{i})
H_{n_k^{i}}(\sqrt{\omega_k}y_{k}^{i})}{2^{n_{k}^{i}}
(n_{k}^{i}!)}W_{kl}^{n^{i}_{k}}
e^{-\frac{1}{2}\omega_k\((x^{i}_{k})^2+(y^{i}_{k})^{2}\)},
\nonumber
\\
\end{eqnarray}
For each value of $i$, $k$ and $l$, the sum over
$n_{k}^{i}$ can be evaluated using the following
identity for Hermite Polynomials \cite{Matur2, Bateman}
\begin{equation}
\sum_{n^{i}_{k}}
\frac{ H_{n^{i}_{k}}(\sqrt{\omega_{k}}x_{k}^{i})
H_{n_{k}^{i}}(\sqrt{\omega_k}y^{i}_{k})}
{2^{n_{k}^{i}}(n_{k}^{i}!)}W_{kl}^{n^{i}_{k}}
=\frac{1}{\sqrt{1-W_{kl}^2}}
\,e^{\omega_{k}\left(
\frac{2x^{i}_{k}y^{i}_{k}W_{kl}^{2}
-\((x^{i}_{k})^{2} +(y^{i}_{k})^2\)W_{kl}^{2}}
{1-W_{kl}^{2}}\right)},
\end{equation}
which gives the result:
\begin{eqnarray}
\Delta\left(\{x^{i}_{k}\},\{y^{i}_{k}\},
\Delta x^{-}, \Delta x^{+}\right )_{\beta}
=
\sum_{l=1}^{\infty}\int\frac{dp^+}{\pi}
e^{-ip^+(\Delta x^{-} -i\frac{\beta l}{\sqrt{2}})}
\nonumber
\\
\times\prod_{i=1}^{D-2}\prod_{k=-\infty}^{\infty}
\left\{\left[\frac{\omega_{k}\,q_{l}^{\omega_{k}}}
{\pi\left(1-q_{l}^{2\omega_{k}}\right)}\right]^{\frac{1}{2}}\,
e^{\frac{\omega_{k}}{1-q_{l}^{2\omega_{k}}}
\left\{2x^{i}_{k}y^{i}_{k}q_{l}^{\omega_{k}}-
\left[\(x^{i}_{k}\)^{2}+\(y^{i}_{k}\)^{2}\right]
\(1+q_{l}^{2\omega_{k}}\)\right\}}\right\}
\label{qexp}
\\
=\sum_{l=1}^{\infty}\int\frac{dp^+}{\pi}
e^{-ip^+(\Delta x^{-} -i\frac{\beta l}{\sqrt{2}})}
\nonumber
\\
\times\prod_{i=1}^{D-2}\prod_{k=-\infty}^{\infty}
\left\{\left[\frac{i\omega_{k}}
{2\pi\sin(\pi\tau_{l}\omega_{k})}\right]^{\frac{1}{2}}\,
e^{\left\{
\frac{i\omega_k}{2\sin(\pi\tau_{l}\omega_{k})}
\left[2x^{i}_{k}y^{i}_{k} 
-\((x^{i}_{k})^2 +(y^{i}_{k})^2\)
\cos(\pi\tau_{l}\omega_k)
\right]\right\}}\right\}
\nonumber
\\
\end{eqnarray}
This is a very nice result, since this expression has
the same structure as the zero temperature propagator
\cite{GS} (see also \cite{chong}). The only difference is
the sum over $l$ and the dependence on $\tau$ defined in
$(\ref{tau})$ instead of just $\Delta{x^+}$. After making
the same manipulation in the zero temperature part, the
expression for the real-time thermal light-cone propagator
for closed pp-wave strings is: 
\begin{eqnarray}
G(\{x^{i}_{k}\},\{y^{i}_{k}\}, \Delta x^{-},\Delta x^+)_{\beta}
=\int \frac{d p^+}{2\pi}e^{-ip^+\Delta x^{-}}G(0,0,0,\Delta x^{+})
\nonumber
\\
\times\prod_{i=1}^{D-2}\prod_{k=-\infty}^{\infty}
e^{\left\{\frac{\omega_{k}}
{2i\sin (\frac{\Delta x^{+}\omega_{k}}{2p^{+}} )}
\left[2x^{i}_{k}y^{i}_{k} - \((x^{i}_{k})^2 +(y^{i}_{k})^2)\)
\cos(\frac{\Delta x^{+}\omega_{k}}{2p^{+}})\right]\right\}}
\nonumber
\\
+ \sum_{l=1}^{\infty}\int\frac{d p^+}{\pi}
e^{-ip^+(\Delta x^{-} -i\frac{\beta l}{\sqrt{2}})}
G(0,0,0,\tau_{l})
\nonumber
\\
\times\prod_{i=1}^{D-2}\prod_{k=-\infty}^{\infty}
e^{\left\{\frac{i\omega_{k}}{2\sin(\pi\tau_{l}\omega_{k})}
\left[2x^{i}_{k}y^{i}_{k}-\((x^{i}_{k})^2 +(y^{i}_{k})^2\)
\cos(\pi\tau_{l}\omega_{k})\right] \right\}},
\label{prod}
\end{eqnarray}
with
\begin{equation}
G(0,0,0,\tau_{l})=\prod_{k=-\infty}^{\infty}
\left[\frac{\omega_{k}\,q_{l}^{\omega_{k}}}
{\pi\left(1-q_{l}^{2\omega_{k}}\right)}\right]^{\frac{(D-2)}{2}}
=
\prod_{k=-\infty}^{\infty}
\left[\frac{i\omega_{k}}
{2\pi\sin(\pi\tau_{l}\omega_{k})}\right]^{\frac{(D-2)}{2}},
\end{equation}
and $G(0,0,0,\Delta x^{+})=G(0,0,0,\tau_{l=0})$.
The expression for the propagator (\ref{prod}) can be written in
terms of modular functions, at least for the flat space limit.
Consider the temperature dependent part written in terms of
$q_{l}$.
Replacing the expression for the Fourier coefficients
$x^{i}_{k}$ and $y^{i}_{k}$, it can be presented as
\begin{eqnarray}
\Delta\left(\{x^{i}_{k}\},\{y^{i}_{k}\},
\Delta x^{-}, \Delta x^{+}\right )_{\beta}= 
\sum_{l=1}^{\infty}\int\frac{dp^+}{\pi}
e^{-ip^+(\Delta x^{-} -i\frac{\beta l}{\sqrt{2}})}
G(0,0,0,\tau_{l})
\nonumber
\\
\times\prod_{i=1}^{D-2}
e^{\sum_{k=0}^{\infty}\left(\frac{\sqrt{2}}{2}\right)^2
\int \frac{d\sigma}{2\pi p^{+}}\frac{d\sigma'}{2\pi p^{+}}
\left\{
\frac{\omega_{k}}{1-q_{l}^{2\omega_{k}}}
\left\{2x^{i}(\sigma)y^{i}(\sigma')q_{l}^{\omega_{k}}-
\left[\(x^{i}(\sigma)\)^{2}+\(y^{i}(\sigma')\)^{2}\right]
\(1+q_{l}^{2\omega_{k}}\)\right\}
\cos\(\frac{k(\sigma-\sigma')}{2p^{+}}\)\right\}}.
\label{tpropags}
\end{eqnarray}
In the flat space limit, $\omega_{k}=|k|$, the
terms in the exponent are manipulated using the
following properties of Theta functions \cite{livro,SIERRA,Tow}
\footnote{The expression (\ref{series}) only works
if $|Im(z)|< \frac{1}{2}Im(\pi\tau)$, that is the case.}:
\begin{equation}
\frac{\Theta_4^{\prime}(z,q)}{\Theta_4(z,q)}
=4\sum_{k=1}^{\infty}\frac{q^k\sin(2kz)}{1-q^{2k}},
\label{series}
\end{equation}
where the prime denotes differentiation with
regard to z, and  $\Theta_4(z,q)$ is defined as
 usual:
\begin{equation}
\Theta_4(z,q)
=\sum_{k=-\infty}^{+\infty}(-)^kq^{k^2}e^{2kiz}.
\end{equation}

Considering
$z=\frac{\sigma-\sigma'}{4p^+}$, and the following results
from (\ref{series})
\begin{eqnarray}
\sum_{k=1}^{\infty}
\frac{2k q^{k}\cos(2kz)}{1-q^{2k}}
&=&\frac{1}{4}\frac{d}{dz}
\frac{\Theta_4^{\prime}(z,q)}{\Theta_4(z,q)}, 
\\
\sum_{k=0}^{\infty}\frac{k(1+q^{2k})}{1-q^{2k}}\cos(2kz)
&=&\frac{1}{8}\frac{d}{dz}\frac{\Theta_{1}^{\prime}(z,q)}
{\Theta_{1}(z,q)},
\end{eqnarray}
the thermal closed string propagator can be elegantly expressed as:
\begin{eqnarray}
&&G^{flat}\left(x^{i}(\sigma),y^{i}(\sigma'),
\Delta x^{-},\Delta x^{+}\right)_{\beta}
= 
\nonumber
\\
&&\int \frac{d p^+}{2\pi}
e^{-ip^+\Delta x^{-}}G(0,0,0,\Delta x^{+})
\nonumber
\\
&&\times\prod_{i=1}^{D-2}
\exp\left[\frac{1}{8}\int \frac{d\sigma d\sigma'}{(2\pi p^{+})^{2}}
\left(x^i(\sigma)y^i(\sigma')\frac{d}{dz}
\frac{\Theta_4^{\prime}(z,q')}{\Theta_4(z,q')} 
- \frac{1}{2}\left(x^i(\sigma)x^i(\sigma') 
+y^i(\sigma)y^i(\sigma')\right)\frac{d}{dz}
\frac{\Theta_{1}^{\prime}(z,q)}
{\Theta_{1}(z,q)}\right)\right]
\nonumber
\\
&&+\sum_{l=1}^{\infty}\int\frac{dp^{+}}{\pi}
e^{-ip^{+}\Delta x^{-}}G(0,0,0,\tau_{l})
\nonumber
\\
&&\times\prod_{i=1}^{D-2}
\exp\left[\frac{1}{8}\int \frac{d\sigma d\sigma'}{(2\pi p^{+})^{2}} 
\left(x^i(\sigma)y^i(\sigma')\frac{d}{dz}
\frac{\Theta_4^{\prime}(z,q_{l})}{\Theta_4(z,q_{l})} 
-\frac{1}{2} \left(x^i(\sigma)x^i(\sigma') 
+y^i(\sigma)y^i(\sigma')\right)\frac{d}{dz}
\frac{\Theta_{1}^{\prime}(z,q_{l})}{\Theta_{1}(z,q_{l})}
\right)\right], \label{teta}\nonumber \\
\end{eqnarray}
where $q=q_{l=0}$. The Theta functions
appearing here reflect the doubly periodicity of
this propagator. In fact, the zero temperature part
can be given in terms of Green functions on a torus
\cite{GS}. The temperature corrections have the same doubly
periodicity with a typical winding. Note that it was not
needed to talk about time compactifications to get this winding.
This is an effect of the Bogoliubov transformation and we are
going to talk more about this in the next section. On the
other hand, it will be interesting to give a path integral
representation for this propagator in terms of a Schwinger's
closed-time path and see how the winding appears in this
scenario.

Now it must be emphasized that the propagators given
at equations $(\ref{prod})$ and $(\ref{teta})$ are not
really the physical propagators because they do not satisfy
the level matching conditions. It is straightforward to take
into account the constraint $(\ref{lmc})$ just because this
propagator has the same structure of zero temperature one.
In general, for the zero temperature propagator, the constraint
$(\ref{lmc})$ can be improved using the Lagrange multiplier
$\lambda$ as follows 
\begin{eqnarray}
G^{phys.}\left(X(\sigma),Y(\sigma')\right)
&=& \int d\lambda\left\langle0\right|
\Phi\left(X(\sigma)\right)
e^{\sum_{i}\sum_{k}2\pi\lambda kN_{n}^{i}}
\Phi\left(Y(\sigma')\right)\left|0\right\rangle 
\nonumber 
\\
&=&\int d\lambda\left\langle 0\right|
\Phi\left(X(\sigma)\right)
\Phi\left(Y(\sigma'+ 2\pi \lambda )\right)
\left|0\right\rangle. 
\end{eqnarray}
As the Bogoliubov generator commutes with the first
quantized operator ${\sum_{n}nN_n^i}$, the same manipulation
can be done for the thermal propagator. So, to get the
physical propagator, it is necessary just to replace 
$Y(\sigma')$ in (\ref{teta}) by $Y(\sigma'+ 2\pi \lambda )$
and integrate over $\lambda$ \cite{Tow}. We can do the same
in the pp-wave propagator (\ref{prod}) after an inverse Fourier
transformation.  

\section{Conclusions and Discussions.}
\label{concdiss}

In this work a finite temperature formulation of pp-wave
light-cone string theory was done using TFD; a real-time
canonical approach. The
main motivation of such an endeavor is to understand the
role of superstring interactions in the Hagedorn behaviour.
This is the core of a research program that has just started
with this paper. The construction of  a finite temperature
string field allows us to  understand the true dominant
string objects at high temperatures.  For example, it was
pointed out in \cite{Greene} that at finite string coupling,
multi-string bound states may become dominant at high energies,
differently from the case with zero string coupling, when the
high energy behaviour is effectively that of a single string.
The LCSFT at finite temperature is  suitable to investigate 
this point, which is very important to really understand what
the Hagedorn temperature means. In addiction, LCSFT at finite
temperature provides a well suited way to show how the
dictionary defined in \cite{BMN} works at finite temperature. 

In this paper all the TFD ingredients were developed in free
light-cone string field. The free energy of an ideal light-cone
pp-wave string gas was derived. Although the expression for this
free energy is already known
\cite{Greene,zayas,semen,sugawara,park,POS,TORUS},
it is the first time that a light-cone string gas is derived from
a second quantized string theory, which is the natural scenario to
study multi-string effects in light-cone. In addition, the TFD
formulation of LCSFT allowed  to derive a real-time thermal
light-cone string propagator, which can be very useful to understand
the dynamics of the theory at finite temperature. 
At this point we would like to make some comments on the results
given in (\ref{sqfree}) and (\ref{prod}). The string gas free energy
is usually derived by
evaluating the partition function on a torus, defined in a target
space with  time coordinate compactified in a circle of radius beta,
which originates a winding sum. In general, this winding provides
a tachyonic spectrum, which in some papers is interpreted as
the source of the Hagedorn divergence. Here, it is necessary
to emphasize that in the TFD approach, it is neither needed to
talk about torus nor explicit time compactification. However,
there is somehow a time periodicity and a torus structure
behind the thermal vacuum defined in (\ref{esqtv}).
This is clear looking at the expression (\ref{teta}) for the thermal
propagator, that has the torus doubly periodicity and a typical
winding sum. To understand this interesting point, let's go back
to the first quantized string in TFD approach. The transverse
partition function for a single string in light-cone is
\begin{equation}
z_{lc}={\mbox Tr}
\left[e^{-\beta h+2\pi i\lambda p}\right],
\label{Z}
\end{equation}
where $\lambda$ is a Lagrange multiplier that imposes the level
matching condition, $h$ is the first quantized Hamiltonian and
$p$ is the  momentum operator that generates translations in
the world sheet $\sigma$ coordinate. 
This expression defines a torus with moduli space parameters
defined by $\tau =\lambda + i\frac{\beta}{2\pi}$.

The same thermodynamic results derived from this partition
function can be obtained in TFD approach, just making the
same manipulations made here in the first quantized context.
In this case, the tilde system is interpreted as an auxiliary
string (tilde string) propagating backwards in Euclidean
time \cite{TORUS}, and the first quantized thermal vacuum has
a clear and beautiful topological interpretation. It was shown
in \cite{POS,TORUS}, that the first quantized thermal vacuum
is a boundary state for the gluing of the string and tilde string,
in order to make a torus with the same moduli space defined
by (\ref{Z}). This gluing is the entanglement produced
by the Bogoliubov transformation, that confines the field in a
restricted region of the time axis. In fact, a generalized
Bogoliubov transformation has been used to describe compactified
bosonic and fermionic fields \cite{adm,admh}. As shown in section
\ref{FTS}, in the second quantized scenario
we have already started with a first quantized string and tilde
string. This is clear in the expansions (\ref{dnexp}). 
In some ways, the effect of the second quantized Bogoluibov
transformation is to produce a torus with these two first
quantized strings plus a time compactification. The key to
understand this effect lies on the thermal state condition
(\ref{sqtsc}). In the first quantized case, the thermal
state condition is precisely the
boundary equation for the gluing of the string and tilde string.
This suggests that the second quantized thermal state condition
may play a role of a boundary state for the gluing of two strings,
propagating in a target space with compactified time coordinate.
This point needs to be better investigated and will help to
understand the TFD structure in second quantized string theory.
Besides this, the flat space thermal propagator derived here is
written in terms of Theta functions plus a typical winding sum.
It will be interesting to show that this propagator can be
derived from the torus Green equations, taking into account the
KMS boundary conditions.

Finally, let's just write a few lines about how the
interactions can be introduced. The main characteristic of
the TFD approach, which really makes it suitable to study
string interactions at finite temperature,  is the use of a
canonical quantum mechanical perturbation theory to calculate
the free energy of an interacting  gas. Following Umezawa
\cite{ume93}, if the interaction Hamiltonian is defined by
$H_I$, one can define the parameter $s$ by  
\begin{equation}
H(s)= H_0 +sH_I,
\end{equation}
where $H_0$ is the free Hamiltonian. In the Heisenberg picture,
the free energy for this interaction system is:
\begin{equation}
F= F(0) +
\int_{0}^1ds \left (0(\theta,s)\right|H_I\left|0(\theta,s)\right)
\end{equation}
where $F(0)$ is the ideal gas free energy derived here and
$\left|0(\theta,s)\right)$ is the thermal vacuum defined for
the Hamiltonian $H(s)$. This expression shows that the variation
of the free energy due to an interaction is proportional to the
interaction thermal energy. There is a similar expression where
the expectation value is taken over the thermal vacuum
$\left|0(\theta)\right)$ in the interaction representation:
\begin{equation}
F= F(0) +
\int_{0}^1ds
\left (0(\theta)\right|\left[TH_I(t_0)
\exp\left(-i\int_\infty^\infty dt s \hat{H}_I(t)\right)\right ]
\left|0(\theta)\right), \label{rep}
\end{equation}
where $\hat{H}_I(t)$ is $H_I - \widetilde{H}_I$  in an arbitrary
time $t$. So, with the string interaction Hamiltonian constructed
in \cite{SPRADLIN1,SPRADLIN2,ARY1,ARY2}, it is possible to define
a vacuum $\left|0(\theta,s)\right)$ or an interaction representation
as in (\ref{rep}), and to calculate the non ideal string free
energy perturbatively. This further development of TFD in LCSFT
context is a work in progress.

\section*{Acknowledgements}

We would like to thanks Dafni Z. Marchioro for useful suggestions.
M. C. B. A. was partially supported by the CNPq Grant
302019/2003-0, A. L. G. and D. L. N. are supported by
FAPESP post-doc fellowships.

\end{document}